\documentclass[]{article}
\usepackage{amssymb}
\begin{document}

\title{Spin-statistics theorem and geometric quantisation}
\author{Charis Anastopoulos \thanks{anastop@phys.uu.nl}, \\
\\Spinoza Instituut, Leuvenlaan 4, \\
3584HE Utrecht, The Netherlands  }
\maketitle

\begin{abstract}
We study the relation of the   spin-statistics theorem  to the
 geometric  structures on phase space, which are introduced in quantisation
procedures (namely a $U(1)$ bundle and connection).
 The relation can be proved in both the relativistic and the non-relativistic
 domain (in fact for any symmetry group including internal symmetries)
by requiring that the exchange can be implemented  smoothly  by a class of symmetry
 transformations  that project in the  phase space  of the joint system  system.  We
 discuss the  interpretation of this requirement,  stressing the fact
that any distinction of identical particles comes solely from the
choice of coordinates - the exchange then arises from suitable
change of coordinate system.
We then examine  our construction in the geometric  and
the coherent-state-path-integral quantisation schemes.
 In the appendix we apply our results
to exotic systems exhibiting
 continuous  ``spin''  and ``fractional statistics''. This gives
 novel and unusual forms of the spin-statistics relation.
\end{abstract}

\renewcommand {\thesection}{\arabic{section}}
 \renewcommand {\theequation}{\thesection. \arabic{equation}}
\let \ssection = \section
\renewcommand{\section}{\setcounter{equation}{0} \ssection}

\section{Introduction}

The relation between spin and statistics is a theorem of
relativistic quantum field theory.  In the original proof of Pauli
\cite{Pauli42}  the
spin-statistics theorem arises as a consequence of: i) the
existence of a representation of the Poincar\'e group, ii)
positivity of energy, iii) the necessity that two fields at
spacelike separation either commute or anticommute.

This particular proof was valid for free fields: for generic field
theories one has to recourse to the axiomatic method: identify a
number of postulates as fundamental for a relativistic quantum
field theory and recover general properties of the fields as
consequence of these axioms. Then indeed,   the spin-statistics
connection is verified as a theorem \cite{StrWight}.

These proofs and their variations assume either the postulate
(iii) we gave earlier or a logically  equivalent form for it (see
\cite{DuSu98} for an analysis). However familiar this postulate
might be, it  is not apparently intuitive in the context of a
quantum theory in which fields are the {\em sole} fundamental
objects. It is true  that commutativity at spacelike separations
is an indication of locality, in the sense that measurements of
commuting quantities can be carried out simultaneously. However,
this is  not true for anticommutativity. Hence, the physical
significance of postulate (iii) arises from the consideration of
the relation between quantum fields and relativistic free
particles with spin. This requires the introduction of the duality
field - particle in our interpretation.

This is one reason, why there has been an effort to prove the
relation between spin and statistics in the context of particle
quantum mechanics, without making any reference to quantum fields.
Another motivation for this effort  comes from  the fact
that non-relativistic quantum mechanics (with the Galilei group as
a group of covariance) is a logically complete theory. For this
reason  it would be desirable to prove the spin-statistics
relation  making reference solely to concepts of this theory.

Many proofs of the spin-statistics theorem have been found in this
context. They typically employ a configuration space
representation for the wave functions. The configuration space for
a single particle is taken as a product of ${\bf R}^3$, for the
translational degrees of freedom, times the two-sphere $S^2$ or
the rotation group $SO(3)$ to account for the spin degrees of
freedom.

However, all such proofs use  axioms or principles, which lie
outside the scope of particle mechanics. Hence, Finkelstein and
Rubinstein  considered particles as extended solitonic objects and
used topological arguments from rubber band twisting \cite{FiRu68,
Tsch89}; Balachandran {\em et al} introduced antiparticles and
symmetries reminiscent of the field theoretic CTP \cite{BS}; Berry
and Robbins employed a particular construction for the transported
basis of the spins (using the Schwinger representation of spins)
\cite{BR,Anan}. All constructions need essentially to satisfy a
condition identified in \cite{Gue}.

In this paper we shall study the manifestation of the spin
statistics relation on the {\em classical phase space} of the
particles rather than the configuration space. The phase space
itself as a symplectic manifold does not have enough structure to
support a spin-statistics theorem; we need to add some additional
structure to move towards quantum theory. In geometric
quantisation \cite{Sour, Wood}
this  a $U(1)$ bundle and a  connection compatible
with the symplectic form.

There are various reasons why we think it is of
interest to see the spin-statistics relation in this perspective:
\\ \\
i) We are of the opinion  that indistinguishability is a
statistical  rather than an intrinsic (or ontological) property of
physical systems. By this we mean that if it is not possible to
distinguish  between two  particles  {\em at all times} by
properties either intrinsic or extrinsic to them, then any
statistical scheme we introduce in order to describe the combined
system has to treat these particles as identical. It makes no
difference, whether the corresponding probability theory is
quantum or classical \footnote{Note, however, that in a
deterministic theory particles are distinguishable by  virtue of
their initial conditions.}. Hence, the study  the spin-statistics
relationship enables us to compare the quantum and classical
notions of indistinguishability, and their consequences, with the
aim to identify their geometric origins. The quantum to classical
transition is also of interest in this context.
 \\ \\
ii) There exists  a  theory of symplectic group actions (e.g. the
Poincar\'e group) in close analogy and with the same degree of
generality as the theory of group representations on a Hilbert
space. The spin degrees of freedom arise naturally by the
consideration of the symplectic actions  of the spacetime symmetry
group and do not have  to be put in by hand. Furthermore, by
stating the spin-statistics relation in a geometric language we
may find such relations in more general systems, than ones that
have so far been studied (such systems  may not involve actual
spin degrees of freedom).
\\ \\
iii) The geometric structure that is responsible for the
non-trivial spin-statistics relation is  known as {\em
prequantisation} of a symplectic manifold. This is present in all
quantisation algorithms either as an object that needs to be
introduced {\em a priori} (geometric quantisation, Klauder's
quantisation \cite{Klau}) or as  a structure that arises {\em a
posteriori} after the quantum theory has been constructed (from
the study of coherent states). It can, therefore,  be argued that
this is the minimal structure one needs to add to classical
mechanics, before starting the construction of quantum theories.
As such, we expect our  results to be relevant to formulations of
quantum theory, which try to sidestep the Hilbert space formalism
(see \cite{CA01, CA03} for our perspective).
\\ \\ \\
In our effort to prove the spin-statistics connection, we find,
like all previous works,   that one  needs to introduce  an
additional postulate. Indeed, we identify a postulate that is
simple from a geometrical perspective and show its equivalence to
the spin-statistics connection. However, it is  equally {\em ad
hoc} as far as the relation with the standard formulation of
quantum mechanics is concerned.

In fact, our study does not need to take into account the full
quantisation algorithm: the spin-statistics connection can be
phrased at the level of prequantisation, {\em i.e.} before
constructing the {\em physical Hilbert space}. This latter
construction can be achieved in different ways through the
introduction of  additional structures: in standard geometric
quantisation a {\em polarisation}, in Klauder's theory a
homogeneous metric by which to support a Wiener process. Our
condition is compatible with such additional structures. However,
it remains equally {\em ad hoc} in virtue of standard quantum
theory. The reason is that the probabilistic/statistical concepts
of quantum theory make only indirect (if at all) reference to the
geometrical objects that were used in their construction.

In the next section we give a brief, but self-contained summary of
the basic ideas of geometric quantisation. We then explain how the
combination of subsystems is effected. The  study of the systems
with $SO(3)$  symmetry is our guide,  in order to identify a
general group-theoretic postulate that is equivalent to the
spin-statistics connection. It is then easy (but rather involved)
to generalise for the  Poincar\'e and Galilei groups, and also to
systems with more exotic spin and statistics structure. This
generalisation is found in the appendix.

Overall, our presentation relies on   Souriau's monograph
\cite{Sour}, to which we refer for a detailed treatment of the
existing  material we have found necessary to include in our
paper. We use a different notation, though.

\section{Geometric quantisation}
\subsection{Prequantisation}
The state space of classical mechanics is a symplectic manifold,
i.e. a manifold $\Gamma$ equipped with a non-degenerate two-form
$\Omega$, which is closed ($d \Omega = 0$). $\Omega$ is known as
the {\em symplectic form}; its physical significance lies in the
fact that it provides a map from observables $f$ (functions on
$\Gamma$) to vector fields $X_f$ (that generate one-parameter
groups of diffeomorphisms) through the assignment
\begin{equation}
d f = \iota_{X_f} \Omega
\end{equation}
(Here $\iota$ denotes the interior product). Vector fields that
can be written as $X_f$ for some $f$ are called {\em Hamiltonian}.
The Poisson bracket between two functions $f$ and $g$ is then
defined as $X_{\{f,g\}} = - [X_f,X_g]$.

A group action on the  manifold is called {\em symplectic} if
its generating vector fields are all Hamiltonian.

Passing from classical to quantum mechanics
 necessitates the introduction of complex-valued
objects. The most natural way to achieve this  is through
 a gauge $U(1)$ symmetry. This  allows us to implement the rule that Poisson
bracket goes to operator commutator.

 More precisely the prequantisation of
a symplectic manifold $(\Gamma, \Omega)$ consists of a fiber
bundle $(Y,\Gamma,\pi)$ with total space $Y$,  base space
$\Gamma$, fiber and structure  group  $U(1)$, with $\pi:Y
\rightarrow M$ the projection map \footnote{One also employs the
associated line bundle $(L,\Gamma,\pi')$, which has ${\bf C}$ as
fiber.}. In addition $Y$ is equipped with a connection, whose form
$\omega$ satisfies $  d \omega  = \pi_* \Omega$.

An immediate consequence of this definition is that one cannot
prequantise all symplectic manifolds:  an integrability condition
arises, which the symplectic form has to satisfy. This comes from
the fact that a connection generates parallel transport along
paths. If $A$ is a potential of a connection $\omega$ , then the
holonomy along a loop $\gamma$ $\exp(i \int_{\gamma}A)$ equals
$\int_{\Sigma} dA = \int_{\Sigma} \Omega $, where $\Sigma$ is a
two-surface   spanning $\gamma$. Since the holonomy is independent
of the choice of $\Sigma$,  $\int_{\Sigma}\Omega$ is an integral
multiple of $2 \pi$ \cite{Wood}.
\\ \\
We need to point out  two facts  that we will use in what follows: \\ \\
i) The inequivalent prequantisations -if any exist- of a symplectic manifold
are classified by the characters of its homotopy group. Hence a simply-connected manifold
has a unique prequantisation. \\
ii) If there exists a symplectic potential (i.e. an one-form $\theta$ on $\Gamma$ such that
$d \theta = \Omega$ globally) the prequantizing bundle is trivial.

\subsection{Group actions}
To any function on $\Gamma$ there corresponds a unique vector
field $Y_f$ on $Y$, such that $\omega(Y_f) = \pi_*f$ and
$\iota_{Y_f}d \omega =  \pi_* df$. Vector fields of the type $Y_f$
generate diffeomorphisms  on $Y$ that are known as {\em
quantomorphisms}.

A group $G$ that acts on $Y$ by symplectomorphisms as $(g \in G,
\xi \in Y) \rightarrow g \cdot \xi$ has also a symplectic action
on $\Gamma$ as $ (g \in G, x \in \Gamma) \rightarrow \pi(g \cdot
\xi)$, for any $\xi$ such that $\pi(\xi) = x$. The action of $G$
on $Y$ is then called a {\em lifting} of the symplectic action of
$G$ on $\Gamma$. Because of topological obstructions not all
symplectic group actions can be isomorphically lifted. Often one
needs introduce a larger group $G'$ acting on $Y$, which is a
covering group of $G$.

We shall denote the defining $U(1)$ action on the bundle as $(z
\in U(1), \xi \in Y) \rightarrow z \cdot \xi$.

\subsection{The next step}
One can construct a Hilbert space from the cross-sections of the
line bundle associated with $Y$. From this,  a natural  assignment of
functions on $\Gamma$ to operators on this Hilbert space follows. The Hilbert space
 is, however, too big compared with the ones of standard quantum
mechanics. One needs to restrict in one of its subspaces.

Standard geometric quantisation proceeds by choosing a
polarisation $P$ , which amounts to choosing a maximal Lagrangian
subspace $P_x$  \footnote{I.e. a subspace in which the symplectic form
vanishes.}  of the complexified tangent space $T_x^{{\bf C}}\Gamma$ at each point $x \in \Gamma$.
  The physical Hilbert space is constructed by all cross-sections
 of the bundle that are constant along the vector fields of the
polarisation. For instance, in the position representation of the
free particle the polarisation is generated by the vector fields
$\frac{\partial}{\partial p^i}$. In general, there is substantial
freedom in the choice of polarisation, but when the system has a
symmetry,it  is preferable to choose a polarisation that is
 preserved by the symplectic action of the corresponding group.

A different way to proceed is to consider the space of
complex-valued functions on $\Gamma$ and then identify a
projection operator onto the physical Hilbert space. In Klauder's
coherent state quantisation the projector is constructed by a path
integral, in which the connection form plays dominant role. An
homogeneous metric on $\Gamma$ is also necessary, in order to
support a Wiener process for the path integral's definition. The
relation between these two types of quantisation is  found in
\cite{Klau2}.

There are other variations of these themes of quantisation schemes
based on geometry. We only want to point out that the justification of
the geometric structures introduced in the quantisation   are
viewed as intermediate steps towards the construction of the
Hilbert space. Once arriving there, all physical interpretation
takes place through the Hilbert space concepts. In particular,
there does not exist an apparent relation between geometric
objects and the statistical ones of standard quantum theory.

\section{Combination of subsystems}
In classical mechanics (or any classical statistical system) the
combination of subsystems is effected through the Cartesian
product. That is, if $(\Gamma_1, \Omega_1)$ and $(\Gamma_2,
\Omega_2)$ are phase spaces associated with two physical systems,
then  the combined system is described by the Cartesian product
$\Gamma_1 \times \Gamma_2$ and the symplectic form $\Omega =
\Omega_1 \oplus \Omega_2$ \footnote{ If $X_1, Y_1$ are vector
fields on $\Gamma_1$ and $X_2,Y_2$ on $\Gamma_2$, then $\Omega_1
\oplus \Omega_2$ is defined by $\Omega_1 \oplus
\Omega_2[(X_1,X_2),(Y_1,Y_2)] = \Omega_1(X_1,Y_1) +
\Omega_2(X_2,Y_2)$. It is similarly defined for al tensor fields.}.

As we mentioned in the introduction, the notion of identical
systems is meaningful also in a classical setting, as it is
essentially a statistical one. Two systems are identical if they
cannot be distinguished at all times by virtue of any internal or
external characteristics.

Even though   two particles can always be distinguished by virtue
of their initial conditions in a deterministic theory, symplectic
geometry is also an arena for statistical description of physical
systems. Symmetries are generated by a Hamiltonian flow, but this
is  the case for  dynamics only if time-translation is a symmetry.
This is not the case in, for instance, open systems.

If we have then two identical systems, any function
 on $\Gamma \times \Gamma$, has to be symmetric with
respect to the exchange

\begin{equation}
(x_1, x_2) \rightarrow (x_2, x_1).
\end{equation}

 The existence of this symmetry amounts to having a probabilistic description
 in terms  of functions on a  phase space $\Gamma_S$. The latter  is obtained
  the  following way. We first define the diagonal set
 $\Delta = \{ (x,x), x  \in \Gamma \}$. Then
$\Gamma_S$ is defined as the quotient of $\Gamma \times \Gamma -
\Delta$ with respect to the permutation (3.1). If $p: \Gamma
\times \Gamma - \Delta \rightarrow \Gamma_S$ is the corresponding
projection map, there exists a unique symplectic form
 $\Omega_S$ on $\Gamma_S$, such that  $p_* \Omega_S = \Omega \oplus \Omega$.

This definition is easily extended for more than a pair of
identical systems. For $n$ systems we define the diagonal $\Delta
= \{ (x_1, x_2, \ldots, x_n) \in \Gamma^n | \exists  i,j , \hspace{0.22cm} s.t. \hspace{0.22cm}
x_i = x_j \}$. The resulting space  $\Gamma_S$ is the quotient of
$\Gamma^n - \Delta$ with respect to the group of permutations.

When    two systems $(Y_1, \Gamma_1, \pi_1; \omega_1)$ and $(Y_2,
\Gamma_2, \pi_2; \omega_2)$ are combined at the prequantisation
level, the total system is described by a fiber bundle with basis
space $\Gamma_1 \times \Gamma_2$ and  a total space $Y$, which is
constructed as follows. We define the 1-form $\tilde{\omega} =
\omega_1 \oplus \omega_2$ on $Y_1 \times Y_2$ and
 then identify the null direction of $\tilde{\omega}$, i.e a vector field $Z$ on
$Y_1 \times Y_2$, such that $\tilde{\omega }(Z) = 0$. This defines
a foliation on $Y_1 \times Y_2$. In fact, a leaf of this foliation
is characterised by the group action
\begin{equation}
(\xi_1, \xi_2) \in Y_1 \times Y_2 \rightarrow (z \cdot \xi_1, z^{-1} \cdot \xi_2) ,
 z \in U(1).
\end{equation}
Hence one can define $Y$
as the quotient of $Y_1 \times Y_2$ by this group action.
The one-form $\tilde{\omega}$ naturally projects into an 1-form $\omega$ on $Y$.
Also the projection map is $\pi: Y \rightarrow \Gamma_1 \times \Gamma_2$ is
defined as $\pi([\xi_1, \xi_2]) =  (\pi_1(\xi_1), \pi_2(\xi_2))$, where
 we denoted as $[\xi_1,\xi_2]$ the equivalence class of $(\xi_1, \xi_2)$ under
  the group action (3.2). The action of $U(1)$ on $Y$ along the fibers is then
  $ z \cdot [\xi_1, \xi_2] := [z \cdot \xi_1, z \cdot \xi_2]$.

Let us consider now the case of identical systems. From a pair of
$\Gamma$ 's one can construct uniquely the bundle $(Y, \Gamma
\times \Gamma, \pi)$. There exists the action of the permutation
group on $\Gamma \times \Gamma - \Delta $, which can be lifted on
$Y$. If $\Gamma$ is simply connected, so is $\Gamma \times \Gamma
- \Delta$ and there are two possible ways by which the permutation
group may act \cite{Bloo80}. Either

\begin{eqnarray}
[\xi_1, \xi_2] \rightarrow [\xi_2, \xi_1], \nonumber
\end{eqnarray}
or \begin{eqnarray}
 [\xi_1, \xi_2] \rightarrow   [(-1)\cdot \xi_2, \xi_1].
\end{eqnarray}

Now if $i : \Gamma \times \Gamma - \Delta \rightarrow \Gamma
\times \Gamma$ is the inclusion map, the actions above also pass
into the pull-back bundle $i_*Y$. From each of these actions we
obtained two different quotient spaces from $i_*Y$ and essentially
to different bundles over $\Gamma_S$ for the prequantisation of
the combined system. They correspond respectively to Bose-Einstein
and Fermi-Dirac statistics and their total spaces will be denoted
as $Y_B$ and $Y_F$ respectively. It is easy to see that the
connection 1-form and the projection maps pass down from $Y$ to
$Y_S$ or $Y_B$.

In fact, the same results are valid for combination of more than
two systems.
  The inequivalent  prequantisations of a connected manifold
are classified by the  characters of its homotopy group.
If $\Gamma$ is connected then the homotopy group
   of  $\Gamma_S$ is the permutation group, which has only
two characters, namely $\chi_+(P) = 1$ and $\chi_-(P) =
\sigma(P)$; here $P$ denotes a permutation and $\sigma(P)$ its
parity.

\section{The spin-statistics relation}
Symplectic geometry has the attractive feature of an intimate
relation with Lie group theory. Given a Lie group, one can
determine all symplectic manifolds,  upon which it acts with
symplectic transformations. They are essentially orbits of the
coadjoint action of the group on the dual of its Lie algebra (for
details see \cite{Sour,Guil}).

This fact provides one of the motivation for studying the
spin-statistics relation in the present context, because spin
degrees of freedom appear naturally from the representation theory
of groups containing space rotations  and they do not have to be
{\em postulated ad hoc}.

For instance, in analogy with Wigner representation theory for the
Poincar\'e group, one can get the symplectic manifolds
corresponding to a free massive or massless relativistic particle
with spin.

\subsection{SO(3) and spin}
The easiest way to understand the appearance of spin classically
is through the study of the symplectic actions of the
group $SO(3)$ of rotations
in 3-dimensional space (ignoring all translational degrees of
freedom). The symplectic manifolds upon which $SO(3)$ acts
transitively have the topology of a two sphere $S^2 =
\{(x_1,x_2,x_3) \in {\bf R}^3; x_1^2 + x_2^2 +x_3^2 = 1 \} $, with
the symplectic form
\begin{equation}
\Omega = \frac{1}{2}s \epsilon_{ijk} x^i dx^j \wedge dx^k
\end{equation}
(It is a different symplectic manifold for each choice of
$s$). The group $SO(3)$ acts as $x^i \rightarrow O^{i}{}_j
x^j$ in terms of its fundamental representation.

The prequantisation of $S^2$ is achieved with the use of spinors.
First, one can show that a necessary and sufficient integrability
condition for $S^2$ to be prequantizable is $s$ to equal $n/2$,
with $n$ an integer, i.e. the usual quantum notion of a spin.

Let us consider a two-spinor
\begin{eqnarray}
\xi = \left( \begin{array}{c} z_1 \\ z_2 \end{array} \right),
\end{eqnarray}
which is normalised to unity $\bar{\xi} \xi = 1$. All such unit
spinors span a 3-sphere $S^3$. There exists a natural projection map
$\pi: S^3 \rightarrow S^2$ given by
\begin{equation}
[\pi(\xi)]^i = \bar{\xi}\sigma^i \xi,
\end{equation}
and a natural connection form
\begin{equation}
\omega = -i \bar{\xi} d \xi,
\end{equation}
and a $U(1)$ action along the fibers $e^{i \phi} \cdot \xi = e^{i
\phi} \xi$. The ensuing bundle is known as the Hopf bundle.

For each choice of $n$ we have the action of the group of $n$-th
roots of unity on $S^3$:
\begin{equation}
e^{2 i  \pi r/n}\cdot \xi, \hspace{0.5cm} r = 0, \ldots, n-1.
\end{equation}
 The prequantisation of a system characterised by a given value of $s$
is a bundle with total  space $Y_n$, which is the space of orbits
$[\xi]_n$ of $S^3$ under the action (4.5). The projection
$\pi_n:Y_n \rightarrow S^2$ is defined as $\pi_n([\xi]_n) =
\pi(\xi)$, while the connection form on $Y_n$ has as pullback on
$S^3$ the connection
\begin{equation}
\omega_n = - i n \bar{\xi} d \xi
\end{equation}
It is easy to check that $ d \omega_n = (\pi_n)_*\Omega$ for the
value $s = n/2$.

Again using the properties of spinors one obtains a lift of the
$SO(3)$ action, or rather of its  double cover $SU(2)$ on $Y_n$. If
$\alpha \in SU(2)$ then to it there corresponds an $SO(3)$ matrix
$O^{ij}(\alpha)  = \frac{1}{2} Tr(\alpha^{\dagger} \sigma^i \alpha \sigma^j)$
 acting or ${\bf R}^3$  and hence on $S^2$. (As is well known, the map
 $\alpha \rightarrow O(\alpha)$ is two-to-one.)  The action of
 $SU(2)$ on $S^3$, which lifts the $SO(3)$ symplectic action on $S^2$, is
$\xi \rightarrow \alpha \xi$. Since this action commutes with the
action (4.5), it passes through the equivalence classes into the
bundle $Y_n$ that prequantises the spin system, with arbitrary
value of $n$.

Due to the equivalence relation coming from (4.5), if $ X$ is an
element of the Lie-algebra of $SU(2)$, the corresponding $SU(2)$
group element for the  action on $Y_n$ is
\begin{equation}
\cos (s |X|) {\bf 1} - \frac{i}{|X|} \sin (s |X|) \sigma^i X_i
\end{equation}
This  shows that a rotation of $|X| = 2 \pi$ performs the transformation
\begin{equation}
\xi \rightarrow (-1)^n \xi,
\end{equation}
 i.e. for even values of $n$ the action of the $SU(2)$ matrix $-1$ is
 identified with the action of unity.

Consider now the combination of two  spin systems with spin $s =
n/2$. According to the general construction presented earlier the
total space for the  bundle characterising the system is an
equivalence class of a pair of unit spinors $[\xi_1, \xi_2]$
modulo the equivalence
relation \\ \\
$(\xi_1, \xi_2) \sim (e^{i\phi} \xi_1, e^{-i \phi} \xi_2)$ and
$(\xi_1, \xi_2) \sim (e^{i2\pi r/n} \xi_1, e^{i2 \pi r'/n} \xi_2)$,
$r,r' = 0, \ldots, n-1$. \\ \\
There is a natural projection to $S^2 \times S^2$ as well as the two
possible actions of the permutation group
\begin{equation}
[\xi_1, \xi_2] \rightarrow [\xi_2, (-1)^f \xi_1],
\end{equation}
We denoted  $f = 0 $ for the bosonic and $f=1$ for the fermionic
action, each value corresponding to the two possible prequantisations of the
combined system.

Clearly, if $G$ is the symmetry group of a phase space $\Gamma$,
then $G \times G$ is a symmetry on $\Gamma \times \Gamma$ and its
action can be lifted on its prequantizing bundle.

The action of a  symmetry group like $SO(3)$ or the Poincar\'e
group can be thought of as corresponding to a change of coordinate
system. In the case of identical systems, it is natural to assume
that the exchange  takes place through a continuous change of
coordinates.

In the spin system the $SO(3)$ group action has as integral curves
of its generators  circles on $S^2$. If we restrict to
transformations generated by one element $X$ of the Lie algebra of
$SO(3)$, we notice that there are two possible routes  from one
point $x_1$ to another $x_2$, corresponding to a smooth path of
$SO(3)$ actions. The reason is that two  points can be connected
by two segments of the circle that is defined by the group action,
say $\gamma$ and $\gamma'$. We take the convention that $\gamma$
starts from $x_1$ and ends at $x_2$ and $\gamma'$ starts at $x_2$
and ends at $x_1$.

Let us denote by $g_1$ and $g_2$ the elements of the group $SU(2)$
that satisfy:
 $(g_1,g_2) \cdot [\xi_1,\xi_2] := [g_1 \cdot \xi_1, g_2
\cdot \xi_2] = (-1)^f[\xi_2,\xi_1]$. One has then that
\begin{equation}
(g_2 g_1) \cdot \xi_1 = (-1)^f \xi_1.
\end{equation}

Since the systems are identical one can represent a state of the
system, by two {\em non-coinciding } points on a single sphere
$S^2$. Let us consider that the exchange
 $g_1 x_1 = x_2$ and $g_2 x_2 = x_1$ is effected by an
 one-parameter group of $SO(3)$ actions for each point, that have a {\em common generator}.
 That is,  we restrict to an exchange that takes place within the orbit
of a single generator, hence both subsystems have to move in the same circle.
 If the first derivative of the map $t \rightarrow
g(t)x$ exists, then the transformation is along an integral curve
of the vector fields generating the $SO(3)$ action on the sphere.

According to our earlier  discussion, our restriction implies that there are two  choices for
 the orbit of the transformation: \\
i.  $x_1 \rightarrow x_2$ through $\gamma$ and $x_2 \rightarrow
x_1$ through $\gamma^{-1}$ (or similarly for
$\gamma'$). \\
ii. $x_1 \rightarrow x_2$ through $\gamma$ and $x_2 \rightarrow
x_1$ through $\gamma'$.
\\ \\
In the first case the orbit of the transformation  necessarily
{\em passes through the diagonal set} and cannot be continuous on
the  phase space $\Gamma_S$ of the combined system. The second case implies that the
action of $g_2 g_1$ on $x_1$ is a rotation of $2 \pi$, hence $(g_2
g_1) \cdot \xi_1 = (-1)^n \xi_1$. Comparing with (4.10) we get
\begin{equation}
 f = n \hspace{0.2cm}mod \hspace{0.2cm} 2,
\end{equation}
  which amounts to the spin-statistics theorem. (The reader may
easily verify that the conclusions remain unchanged, if we allow
rotations of more than $2 \pi$ along the circle: unless $g_2g_1$
is a rotation of an odd number times $2 \pi$ the diagonal set is
crossed by the transformation.)

The generalisation to systems of $n$ particles is immediate as any
permutation of $n$ objects can be written as a product of
exchanges each involving  two of them.

\subsection{ Generalisation}

Note that  $G \times G $
 acts on $\Gamma \times \Gamma$, but its action does not descend on
  $\Gamma_S$, because  the diagonal is not preserved.
  This action  can be decomposed into one  of the type $(x_1,x_2)
   \rightarrow (gx_1, g x_2)$  and one of the type
$(x_1,x_2) \rightarrow (gx_1, g^{-1}x_2)$. The latter  type does
not preserve the diagonal, while the former generically does. It,
therefore, descends into an action of $G$ on $\Gamma_S$,  the {\em
diagonal action}. It is clear from our previous discussion that
the exchange takes place along the orbits of one-parameter
subgroups, which correspond to the diagonal action of $G$ on
$\Gamma_S$.

This allows us to identify a postulate, which leads to an analogue
of the spin-statistics connection for any quantum mechanical
systems, which is characterised by the transitive symplectic
action of a group $G$ on the classical state space. In other
words, this postulate refers to {\em elementary systems}
associated to the group $G$.
  \\ \\
{\bf Postulate 1:} {\em In  combination of two identical systems,
each characterised by a symmetry group $G$,   it should be
possible to obtain the permutation (3.3)  by smooth   transformations along the orbits of the diagonal
action of $G$ on $\Gamma_S$.}
\\
\\
This statement can be made explicit  as follows: \\ \\
We assume that the Lie group $G$ acts transitively by
symplectomorphisms on $\Gamma$. Then   we demand that there should
exist two elements $Z_1$ and $  Z_2$
of the Lie algebra of $G$, each a scalar multiple of the other,   such that \\ \\
i) one can define the paths $ (t \in [0,1], (x_1,x_2)) \rightarrow
(e^{Z_1t}x_1, e^{ Z_2 t}x_2)$, on $\Gamma \times \Gamma $ that do not cross the diagonal. \\
 ii)  if $g_1 = e^{Z_1}$, $g_2 =
e^{ Z_2}$, then   $g_1 x_1 = x_2 $ and
$g_2 x_2 = x_1$. \\
iii) in the lift in the bundle $Y$ we should have $[g_1 \cdot
\xi_1, g_2 \cdot \xi_2] = e^{i \theta} \cdot [\xi_2,  \xi_1]$,
where $e^{i \theta}$ is the phase associated with the exchange.
\\ \\
Then we expect that the spin-statistics relation  arises as a
consequence.
\\ \\
It is important to emphasise that our postulate singles one
particular class of paths by which the exchange  should be
performed: these are  the {\em orbits of one-parameter subgroups
of G}, in its diagonal action on $\Gamma_S$.

\paragraph{Zero spin} Our postulate  suffices to establish the spin-statistics
connection  for  single spins. Moreover, it is compatible with the
bosonic character of the spin zero particles. In the relativistic
case the phase space is ${\bf R}^6$. It is parametrised by a
4-vector $x $ for a fixed value of $x^0$ (say $x^0 = 0$)
 and
a unit timelike vector $I$ \footnote{A more covariant way to
construct it is by the unit timelike vector $I$ and the
equivalence class of spacetime points, where $x \sim x' $ if $x -
x'$ is parallel to $I$.}. Then the symplectic form reads
\begin{equation}
\Omega = m d x^{\mu} \wedge d I_{\mu} = dp^i \wedge dx_i,
\end{equation}
where $p^i = m I^i$ and the Poincar\'e groups acts as $x
\rightarrow \Lambda x +C$, $I \rightarrow \Lambda I$. Here
$\Lambda \in SO(3,1)$ and $C \in {\bf R}^4$.

The prequantisation proceeds by constructing a trivial bundle $R^6
\times S^1$, with elements $(x,I,e^{i \phi})$. The connection form
is $ \omega = p^i dx_i + d \phi$. The action of the Poincar\'e
group lifts  then $(x, I, e^{i \phi}) \rightarrow (\Lambda x +C,
\Lambda I, e^{i \phi})$, i.e. it is trivial on the fibers. This
implies that the only possible choice of prequantisation for
combined systems is the bosonic one, because there does not exist
any symmetry transformation that could reproduce the fermionic
action of the permutation group.

\paragraph{Relativistic particles:}In  appendix A we demonstrate that  postulate 1  provides the
correct spin statistic relation also for the case of relativistic
and non-relativistic particles {\em with spin}. The proof involves
no  concepts other than  the ones we used in  the case of a single
spin, but is provided in some detail for reasons of completeness.

\paragraph{Non-relativistic particles:} For
non-relativistic systems one can obtain the description of spin by
studying the symplectic actions  of the Galilei  group. The phase
space is a product of the sphere and ${\bf R}^6$, which we have
already studied. The symmetry group,  however, does not factorise
into a piece acting on ${\bf R}^6$ and one acting on $S^2$.
However, it is easy for the reader to verify
 that our postulate 1 reproduces the spin-statistics connection, with
   reference now to the Galilei group, rather than the Poincar\'e.

\paragraph{Non-trivial systems:}  In effect, the spin-statistics theorem in the familiar
setting of particles in three spatial  dimensions is equivalent to
the statement that {\em a rotation by $2 \pi$ of a single particle
is physically identical with an exchange} \cite{Gue}.
 Postulate 1 reproduces this fact in a general group-theoretical
 language, thereby providing a generalisation that can be used in a wider class of systems.
 We demonstrate this in Appendix B. There
 we study  the case of the relativistic particle in three dimensions,
 where rather surprisingly only bosonic statistics seem to be acceptable. Also we study a simple example
 for combination of systems with non-simply connected phase spaces. Such systems employ non-trivial
 prequantizing bundles, hence they have more alternatives than Bose-Fermi for statistics and provide novel versions of
 the spin-statistics theorem.

\subsection{After prequantisation} We proceed to study the
possible consequences of the spin-statistics relation, with
respect to the remaining part of the geometric quantisation
procedure.

\subsubsection{Wave functions} Standard
geometric quantisation proceeds by specifying a complex
polarisation $P$ on the phase space of the system. Let us denote
by $\Xi_P(\Gamma)$ the space of vector fields, such that at each
$x \in M$ the corresponding tangent vector lies in the
polarisation. Let us also consider the line bundle $(B, \Gamma,
\tilde{\pi})$, which is associated to the bundle  $(Y, \Gamma,
\pi)$ of the prequantisation of $\Gamma$. This bundle has total
space $B = Y \times {\bf R}^+$, projection map $\pi(\xi, r) =
\pi(\xi)$ and $U(1)$ action $ e^{i \phi} \cdot (\xi, r) = (e^{i
\phi} \cdot \xi, r)$, for $\xi \in Y$ and $r \in {\bf R}^+$.

The connection form $\omega$  induces a covariant derivative
$\nabla$  on the cross-sections of $B$. A cross-section $\psi$ of
$B$ corresponds to a quantum mechanical wave function if $\nabla_X
\psi = 0$ for all vector fields $X \in \Xi_P(M)$, i.e. if the
cross-section vanishes in the directions of the polarisation.

Clearly in a system of two identical particles on $\Gamma$, the
wave functions are  cross-sections $\psi(x_1,x_2)$ of a bundle
over
 $\Gamma \times \Gamma$. Assuming that  $|\psi|^2(x_1,x_2)$
 remains invariant from the exchange \footnote{ The exchange is inherited from the principal bundle $Y$
so the value of $r = \sqrt{\bar{\psi}\psi}$ is not affected.}, and
that both $\Gamma$ carry
 the same polarisation, we obtain the two possible
 behaviors for $\psi$ according to the choice of the action of
 the permutation (3.3)
 \begin{equation}
\psi(x_2, x_1) = (-1)^f \psi(x_1,x_2)
 \end{equation}
A {\em continuous} cross-section $\psi$ satisfying (4.13) for
either choice of $f$ can be viewed as as a cross-section of either
of the bundles $Y_B$ or $Y_F$ defined earlier.

 Given a group action on  $\Gamma$, one can often construct
 polarisations that are left invariant under the
symplectomorphisms by which the group acts; this is true, for instance,
for the Poincar\'e group. Therefore, if $g(t)$ is an one-parameter
group of transformations of such a group, then the transformation
\begin{equation}
\psi \rightarrow g(t) \cdot \psi (x) = \psi(g^{-1}(t)x),
\hspace{1cm} x \in \Gamma
\end{equation}
can be defined acting on the wave functions. A  smooth
cross-section will remain smooth, whenever the $g(t)$ is smooth, as we have demanded.

One can then define the  action of $G \times G$ on the wave
functions of a combined system (on $\Gamma \times \Gamma$)
\begin{equation}
\psi(x_1, x_2) \rightarrow \psi(g_1(t)x_1, g_2(t) x_2).
\end{equation}
This action descends into the  action on wave
 functions on $\Gamma_S$ (of either the bosonic or fermionic type)
 if it preserves the diagonal
 i.e. if for $x_1 \neq x_2$ , $g_1(t) x_1 \neq g_2(t)x_2$ for all $t \in [0,1]$.
In that case one can write the law
\begin{equation}
\psi(x_1,x_2) \rightarrow \psi(e^{Zt_1}x_1, e^{Z t_2} x_2),
\end{equation}
where  $Z$ is an element of the Lie algebra of $G$ and  $t_1,t_2 \in {\bf R}$.
\\
In light of these remarks, postulate 1 can be rephrased as
\\ \\
{\bf Postulate 1a:} One can perform the exchange (4.13) by means
of a smooth transformation of the type (4.16) acting on wave
functions defined on $\Gamma_S$.

\subsubsection{Path integrals} In Klauder's coherent state quantisation,  one
considers the Hilbert space of complex valued functions
 on the phase space $\Gamma$ and identifies a relevant physical subspace by means of a
 projection operator, which is defined by a positive hermitian kernel
  $K(x_1|x_2), x_1,x_2 \in \Gamma$.
A wave function $\Psi$ on the physical Hilbert space is
   a function on
    $\Gamma$ that can be written in the form
\begin{equation}
\Psi_{\alpha_l,x'_l}(x) = \sum_{l} \alpha_l K(x | x'_l)
\end{equation}
Such functions are parametrised by  a finite number of complex
numbers $\alpha_l$ and points $x'_l \in \Gamma$. A group of
transformations
 on phase space amounts to a transformation $K(x|x') \rightarrow
 K(gx|gx')$, which can be immediately translated
in terms of  wave functions.

The kernel $K$ is constructed by path integration as
\begin{equation}
K(x_1 | x_2) = \lim_{\nu \rightarrow \infty} \int Dx(\cdot) e^{i \int_{x(\cdot)} A
- \frac{1}{2 \nu} \int dt h(\dot{x} \dot{x}) },
\end{equation}
where $x(\cdot)$ is a path on phase space in the time interval $[0,T]$, $A$ is a $U(1)$
 potential  one-form on $\Gamma$ (its pullback by $\pi$ on the bundle $Y$ is $\omega$),
 $h$ is a homogeneous Riemannian metric on $\Gamma$ and  the path integration refers
 to the Wiener measure as supported by the metric $g$ and constrained
 by $x(0) = x_1$, $x(T) = x_2$.
 This quantisation scheme, then,  introduces a homogeneous metric in
 order to arrive at the physical Hilbert space.

A manifold upon which a Lie group $G$ acts transitively is a
homogeneous
 space. Namely, there exists a metric that accepts $G$ as a
 group of isometries and the integral curves of the group action correspond to
  geodesics.  Hence, a transformation
$x_1 \rightarrow gx_1, x_2 \rightarrow g x_2$ corresponds to
diffeomorphisms, which leave the connection form and the metric
invariant.

In the combination of two systems the  metric goes to $h_1 \oplus
h_2$. When the systems are identical, there exists a metric on $h$
such that its pullback on $\Gamma \times \Gamma$ equals $h \oplus
h$. The holonomy is  $ \exp( i\int_{(x_1(\cdot),x_2(\cdot))}A_S)$,
with $A_S$ a potential corresponding to the $U(1)$ connection over
$\Gamma_S$. Under the exchange the potentials transform according
to the structure of the bundle over $\Gamma_S$. Effectively, the
holonomies transform as (3.3) and since the  term with the metric
remains invariant under the exchange, we obtain
\begin{equation}
K(x_1,x'_1| x_2, x'_2) = (-1)^f K(x'_1,x_1|x'_2,x_2)
\end{equation}
This leads, through (4.17), to equation (4.13) for the wave
function.

If $G$ is a symmetry group of $\Gamma$ then
 on the kernels for the quantum theory on $\Gamma \times \Gamma$
  there exists the action of the symmetry group $G \times G$ as
\begin{equation}
K(x_1,x_2|x'_1,x'_2) \rightarrow K(g_1x_1,g_2x_2|g_1x'_1,g_2 x'_2)
\end{equation}
Again, it is not always projected on kernels defined on $\Gamma_S$
as it does not preserve the diagonal. However, the diagonal action is preserved. It is important to
stress that
the  corresponding vector fields on $\Gamma_S$ {\em generate isometries of the metric} on $\Gamma_S$.
The exchange would then not affect the Wiener measure for the process on $\Gamma_S$.

Postulate 1 is equivalent to  the following one \\ \\
{\bf Postulate 1b:} The exchange (4.19) can be performed by a
transformation along the integral curves of the diagonal action of
$G$ on $\Gamma_S$.

\subsection{Internal degrees of freedom:} We would like to consider  the spin-statistics
connection for degrees of freedom that correspond to internal
symmetries (isospin, flavour, ...). One possible procedure would
be to introduce a symplectic manifold $\Gamma_{int}$ for these
degrees of freedom. Typically $\Gamma_{int}$ would be a coadjoint
orbit of the corresponding symmetry group (SO(3), SU(3), etc).

However, this procedure does not work, because the internal
degrees of freedom are genuinely {\em discrete}: isospin in
nuclear physics takes only the value ``proton'' or the value
``neutron'': there is absolutely no physical meaning to a
classical symplectic manifold underlying isospin or flavor. The
internal symmetry label the type of particles: in a
first-quantised version these symmetries can only act on the
indices of the wave function; in a second quantised version on the
indices of the fields. They do not enter the process of
quantisation.

The correct way is to consider that the different values of the
internal degrees of freedom correspond to different copies of the
particle phase space. Hence, in the description of isospin we
 have an elementary phase space consisting of $\Gamma_0 =
\Gamma^{(n)} \times \Gamma^{(p)}$. Now $\Gamma^{(n)}$ is
isomorphic to $\Gamma^{(p)}$, however, the presence of the
internal characterisation as neutron or proton does not render the
particles identical: so the physical phase space that describes a
neutron and a proton should not be quotiened out by the
permutation group. The same would hold for the corresponding
prequantising bundles.

An exchange of isospin  corresponds to a map $Ex: \Gamma_0
\rightarrow \Gamma_0$, such that for $(x,y) \in \Gamma^{(n)}
\times \Gamma^{(p)}$, we have $Ex[(x,y)] = (y,x)$. Clearly, such a
transformation cannot be effected by the Cartesian product of
Poincar\'e groups that is the symmetry group of $\Gamma_0$.

However, if  the physical state of the system is invariant under
$Ex$,   the neutron and the proton are identified. Hence, the
correct physical space  is $\Gamma_S$ and the corresponding
prequantising bundle $Y_B$ or $Y_F$. The sign $(-1)^f$
characterising  the bundle is then interpreted in terms of isospin
exchange. Hence, our result $f = n mod 2$ would be interpreted as
saying that the particles with half-integer spin produce a phase
of $(-1)$ in the isospin exchange.

In other words, we can  either say that we have two identical
particles, or that we have two particles characterised by internal
quantum numbers, the exchange of which is a symmetry of the
physical description. In absence of interactions that break the
internal symmetry, there is no way to distinguish between
particles with internal quantum numbers (a neutron is
distinguished from a proton by virtue of the electromagnetic
interaction): hence, at the fundamental level the spin-statistics
theorem for internal degrees of freedom is tautological with the
spin-statistics theorem for identical particles.

\section{Conclusions}
Souriau in his monograph notes that ``... {\em geometry does not provide the relation between
spin and the character $\chi$} (of the permutation) {\em as suggested by experiments...}''.
We showed that, on the contrary, there is a simple  geometric postulate that leads to this relation, but
the question remains at the level of the relevance of the geometric description to the basic
statistical principles of quantum theory.

 There are two points one can make in relation
to our result.

First,  a transitive group acting on phase space may be viewed
passively as corresponding to coordinate changes: this is
definitely true for the Poincar\'e group. The demand then that the
exchange should be implementable with a smooth group action might
be said to correspond to a statement that the distinction between
two particles done in basis of their coordinates is arbitrary and
one should be able to exchange them in the statistical description
by means of a change of reference frame. Even though our results
at present cannot yet fully
 ascertain this  statement,
 they  definitely assert the relevance of the action of the symmetry
group to the existence of a spin-statistics theorem.

The requirement that the coordinate transformation proceeds
continuously (indeed smoothly) is also of interest. More so,
because the description of identical systems on $\Gamma_S$ is
specially relevant when we consider continuous wave functions. The
Hilbert space formulation focuses on the measurability properties
of the quantum state (over the spectrum of any self-adjoint
operator) - similarly as in classical probability
 Our results might be taken as a hint that continuity
  (or smoothness) over the phase space
is an  important ingredient of quantum probability.

In any case, we have showed the spin-statistics connection
 by means of geometric structures over the phase space, while
  employing the statistical notion of indistinguishability.
   Nowhere, was there any need to employ Hilbert space concepts.
 Indeed, the whole analysis is consistent with formulations of
 quantum theory phrased solely in terms of geometrical
 objects -even classes of hidden variable theories. This
 latter point has indeed been an underlying motivation for this
 work. In light of our results, it is important to find a relation between our geometric
description of the spin-statistics theorem and the standard one
that relies on the positivity of  the Hamiltonian of the
corresponding field. This necessitates a geometric understanding
of the procedure of ``second quantisation'', which is a focus of
our present research.

\section*{Aknowledgements} This research was supported through a European
 Community Marie Curie Fellowship with contract number HPMF-CT-2000-0817.

\begin{appendix}
\section{ Relativistic particles} The symplectic
actions of the Poincar\'e group are classified in complete analogy
with Wigner's classification of Hilbert space
 representations. If $M^{\mu \nu}$ and $P^{\mu}$ are the generators of the
 Lorentz and translation group respectively, we can define the Pauli-Lubanski
 four-vector
\begin{equation}
W^{\mu}  = \frac{1}{2}\epsilon^{\mu \nu \rho \sigma}  P_{\nu}
M_{\rho \sigma}.
\end{equation}
Then the mass $m$ defined by $P_{\mu} P^{\mu} = m^2$, and the spin
$s$ defined by $W^{\mu} W_{\mu} = - s^2 m^2$ are invariants of the
action.

The case $s=0$ was explained in section 4.2.

\subsection{$s \neq 0, m \neq 0$} The phase space is ${\bf R}^6
\times S^2$. It is parametrised by a 4-vector $x$ for a fixed
value of $x^0$, by a unit timelike vector $I$ (corresponding as
earlier to 4-momentum) and a unit spacelike vector $J$
(corresponding to a normalised Pauli-Lubanski vector), such that
$I_{\mu} J^{\mu} = 0$. Note that for a fixed value of $I$, $J$
takes value in a two-sphere (one can readily check that for the
case $I = (1, {\bf 0})$).

The symplectic form is \cite{Sour}
\begin{equation}
\Omega = m dx^{\mu} \wedge dI_{\mu} +\frac{s}{2} \epsilon_{\mu \nu
\rho \sigma} I^{\mu}J^{\nu}(dI^{\rho} \wedge d I^{\sigma} - d
J^{\rho}\wedge dJ^{\sigma}),
\end{equation}
with the Poincar\'e group acting as $(x, I, J) \rightarrow
(\Lambda x + C, \Lambda I, \Lambda J)$.

The prequantisation can be achieved with the use of the Dirac
spinors. We remind that a Dirac spinor consists of a pair of
two-spinors  as
\begin{eqnarray}
\psi = \left( \begin{array}{c} \xi_1  \\ \bar{\xi}_2 \end{array}
\right),
\end{eqnarray}
 while the $\gamma$ matrices are defined as
\begin{eqnarray}
\gamma^{\mu} = \left( \begin{array}{cc} 0 &\tilde{\sigma}^{\mu}  \\
\sigma^{\mu}& 0
\end{array} \right),
\end{eqnarray}
where $\sigma^{\mu} = (1,\sigma^i)$ and $\tilde{\sigma}^{\mu} =
(1, -\sigma^{i})$.

Again, it can be shown that one has to restrict to the choice of
$s = n/2$. The explicit construction of the prequantisation is as
follows:

Consider the space $Y$ which is the Cartesian product of ${\bf
R}^3$ (in which the spatial variables $x^i$ live) times a manifold
$S$, which consists of Dirac spinors
 $\psi$ which satisfy the equation
\begin{equation}
\bar{\psi} \psi = 1 \hspace{4cm} \bar{\psi} \gamma_5 \psi = 0.
\end{equation}
As usually we have denoted $\bar{\psi} = \psi^{\dagger} \gamma^0$
and $\gamma_5 = i \gamma^0 \gamma^1 \gamma^2 \gamma^3 $.

There exists a projection map $\pi (x^i, \psi) = ( x^i, I^{\mu} =
\bar{\psi} \gamma^{\mu} \psi, J^{\mu} = \bar{\psi} \gamma^{\mu}
\gamma_5 \psi)$, while the $U(1)$ action up the fibers is
\begin{equation}
e^{i \phi} \cdot (x, \psi) = (x, e^{i \phi} \psi)
\end{equation}
Thus a $U(1)$ bundle $(Y, {\bf R}^6 \times S^2,  \pi)$  is
constructed.

 As in the case of a single spin, we have the action of the group
 of $n$-th roots of the unity on the fibers
\begin{equation}
 e^{i 2 \psi r/n} \cdot (x, \psi) , \hspace{2cm} r = 0, \ldots,
 n-1
\end{equation}
  Taking the quotient
 of this action we can obtain a bundle $(Y_n, {\bf R}^6 \times
 S^2, \pi_n)$, which has as elements equivalence classes $[x,
 \psi]_n$ and projection map $\pi_n([x, \psi]_n ) = \pi(x,\psi)$.
 Upon this bundle one can a connection form, whose pullback on $Y$ is
 \begin{equation}
\omega_n = -i n \bar{\psi} d \psi -  m' I_{\mu} dx^{\mu},
\end{equation}
where $m' = n m$.

One can lift the action of the Lorentz group $SO(3,1)$ to the one
of its double cover $SL(2,{\bf C})$.  An element $\alpha \in
SL(2,{\bf C})$ acts on Dirac spinors by means of the matrix
\begin{equation}
 U(\alpha) = \left( \begin{array}{cc} \alpha & 0 \\ 0&
 \alpha^{\dagger}
\end{array} \right)
\end{equation}
By virtue of the map $ \psi \rightarrow \bar{\psi} \gamma^{\mu}
\psi$, we verify  that for each $\alpha$ there corresponds an
element $\Lambda(\alpha)$ of $SO(3,1)$ such that
\begin{equation}
\Lambda^{\mu\nu}(\alpha) = \frac{1}{4} Tr \left( \bar{U}
\gamma^{\mu} U \gamma^{\nu} \right) = \frac{1}{2} Tr(
\alpha^{\dagger} \sigma^{(\nu} \alpha \tilde{\sigma}^{\mu)}) ,
\end{equation}
where $\bar{U} = \gamma^0 U^{\dagger} \gamma^0$. The map is
two-to-one as $\pm \alpha$ go to the same $SO(3,1)$ element.

The action then of $SL(2,{\bf C}) \ltimes {\bf R}^4$ on $Y_n$ is
\begin{equation}
[x, \psi ]_n \rightarrow [ \Lambda(\alpha) x + C,  \alpha \psi]_n.
\end{equation}
Note that we have written as $[x, \psi]$ the equivalence class of
elements of $Y$ modulo the action (4.19). We should note again
that for even values of $n$ the action of $-1 \in SL(2,{\bf C})$
is identified - due to (4.19)- with the one of unity. Hence for a
rotation of $2 \pi$ we have
\begin{equation}
(x ,\psi) \rightarrow (x, (-1)^n \psi)
\end{equation}

 We are at the position now to consider the
combination of identical relativistic particles say $(x_1, I_1,
J_1)$ and $(x_2, I_2, J_2)$. Any two points $x_1$, $x_2$ can be
identified by a space translation; we can, therefore, focus on the
$I$ and $J$ degrees of freedom and the corresponding action of the
Lorentz group. We can choose a coordinate system such as $I_1 =
(1,{\bf 0})$, $J_1 = (0, {\bf n})$, with ${\bf n} \in S^2$.
Consider now the special case that $I_2 = I_1$ and $J_2 = (0, {\bf
n}_2)$, the group
 actions reducing to the ones of $SO(3)$. The analysis of section 4.1
  passes unchanged in this case. In particular, we can
  write the path from $ {\bf n}_1$ to ${\bf n}_2$
   as $e^{Z_1t} {\bf n}_1$ and from ${\bf n}_2$
  to ${\bf n}_1$ as $e^{Z_2t} {\bf n}_2$, $Z_1$
   and $Z_2$ being elements of
the Lie algebra of $SO(3)$ and $t \in [0,1]$.  According to our
previous analysis, if motion takes place such that the diagonal
set is not crossed
 $e^{Z_1} e^{Z_2}$ corresponds to a rotation of $2 \pi$.

Now we keep $I_1, J_1$ fixed in its previous value,  but we
consider
 a generic  value of $I_2,J_2$ by the action of a Lorentz transformation
 $\Lambda$ on $I_2 = (1, {\bf 0}, J_2 = (0, {\bf n}_2$) we considered earlier.
 The corresponding paths will transform under the adjoint
  group action $ \Lambda e^{Z_1t} \Lambda^{-1}$ and $\Lambda e^{Z_2t} \Lambda^{-1}$.
   However, the whole analysis remains identical; if two paths intersect
   then this property is preserved by the adjoint action. In particular
   $(\Lambda e^{Z_1} \Lambda^{-1} )  (\Lambda e^{Z_1} \Lambda^{-1} )  $
    still corresponds to a rotation of $2 \pi$ and hence an action on the
     bundle as (4.8). The spin-statistics connection then follows.

\subsection{$ m = 0$, $ s \neq 0$} This case can be proven in a
similar fashion to the previous one. The construction of the phase
space is more intricate, though. For massless
 particles the Pauli-Lubanski vector is parallel to the momentum four-vector,
 which is null.  The state of the system is, then, more conveniently
 specified by the use of a spacetime point $x^{\mu}$, a
 null vector $I$ corresponding to the momentum four-vector
  and another null vector $J$, such that $I_{\mu} J^{\mu} = -1$. Let
   us denote the space consisting of the triple $(I,J,x)$ as  $M$.

We then construct the closed two-form
\begin{equation}
\Omega = - \chi s \epsilon_{\mu \nu \rho \sigma} I^{\mu} J^{ \nu}
dI^{\rho} \wedge d J^{\sigma} + dX^{\mu} \wedge d I_{\mu},
\end{equation}
where $\chi = \pm 1$ is the {\em helicity } of the particle. The
Poincar\'e group acts as follows as $(x,I,L) = (\Lambda x + C,
\Lambda I, \Lambda J)$. Note that from the pair $I,J$ one can
define the vectors
  $K_{\mp} = \frac{1}{\sqrt{2}} (I \mp J)$, which are unit
   timelike and spacelike respectively and satisfy
   $K_- \cdot K_+ = 0$.

 We can repeat a similar procedure as in the massive case, in order to get a
 prequantisationon a bundle $Y$
 together with an action of the $SL(2{\bf C}) \ltimes {\bf
R}^4$ on $Y$, with the property that a rotation of $2 \pi$
corresponds to $(x, \psi)
 \rightarrow (x, (-1)^n \psi)$ \cite{Sour}.

Concerning the combination of subsystems, one can repeat without a
change the analysis of the last two paragraphs of A.1; only now it
has to
 make reference to  the vectors $K_-$ and $K_+$, which we defined earlier.
 This establishes the spin-statistics connection
 for the massless relativistic particles. The only difference is
 that the phase space is topologically ${\bf R}^4 \times S^{2}$,
 because the symplectic form (A.13) has a null direction
 corresponding to the vector fields

\begin{equation}
\alpha^{\sigma} \left( \chi s \frac{\partial}{\partial x^{\sigma}}
+ \epsilon^{\mu \nu \rho \sigma} I_{\nu} J_{\rho}
\frac{\partial}{\partial J_{\mu}} \right)
\end{equation}
where $\alpha^{\mu}$ is a four-vector that satisfies
$\alpha_{\mu}I^{\mu} = 0$. These null directions have to be
excised if we are to construct the physical phase space for the
massless particles.

\section{Topologically non-trivial systems}

\subsection{Three dimensions}
An important feature of the phase space analysis is that it can be
phrased with respect to any symmetry group: it is not just
restricted to groups associated with change of reference frames in
four-dimensional Minkowski spacetime. For instance, it is
meaningful for the study of systems in three dimensional Minkowski
spacetime. The symmetry group there is $SO(2,1) \ltimes {\bf R}^3$
with $SO(2)$ being its subgroup  that generates spatial rotations.
The phase space of the relativistic particle consists of $x^{\mu}$
at constant time $x^0$ and a unit timelike vector $I^{\mu}$. The
symplectic form depends on two parameters: the mass $m$ and the
spin $s$.  However, the  topology is ${\bf R}^4$, hence there
exists no sphere corresponding to the spin degrees of freedom.
Rather ``spin'' arises out of a non-trivial symplectic two-form
\begin{equation}
\Omega = m dx^{\mu} \wedge dI_{\mu} + s m \epsilon_{\mu \nu \rho
}I^{\mu} dI^{ \nu}\wedge dI^{\rho}
\end{equation}
However in ${\bf R}^4$ all closed two-forms are exact, the phase
space has a global symplectic potential and hence the
prequantizing bundle is trivial. The spin $s$ then takes all real values in the quantum theory.

Also in complete correspondence  with the system in 4.3.1 the
SO(2,1) group  acts trivially on the fibers. Our postulate then
implies that in three dimensions  {\em the only possible
statistics are the Bose statistics}.

We should note that Bose statistics is the only possibility in all systems with a simply-connected
phase space that have a global symplectic potential. This includes cotangent bundles over simply connected
configuration spaces.

\subsection{Fractional statistics}
It is important to remark that there is no alternative to the
Bose-Fermi statistics if the phase space of the single particle is
simply connected. However, for a connected, non-simply connected
state spaces $\Gamma$, the possible prequantisations are
classified by the characters $\chi$ of the homotopy group
$\pi_1(\Gamma)$, that is all homomorphisms from $\pi(\Gamma)$ to
$U(1)$.

This comes from the fact that  if $\tilde{\Gamma}$ is the universal
cover of $\Gamma$, there exists by definition an action of
$\pi(\Gamma)$ on $\tilde{\Gamma}$, such that its space of orbits
is $\Gamma$. If $\tilde{\Gamma}$ admits a prequantisation to a
bundle $(\tilde{Y}, \Gamma, \pi)$ and the action of $\pi(\Gamma)$
can be lifted then one can quotient as in section 4.3.2 to obtain
the prequantizing bundle  for $\Gamma$. If $(g \in \pi_1(\Gamma),
\xi \in \tilde{Y}) \rightarrow g \cdot \xi$ denotes an action of
$\pi_1(\Gamma)$ in the bundle, then also $(g \in \pi_1(\Gamma),
\xi \in \tilde{Y}) \rightarrow \chi(g) g \cdot \xi$ is an
inequivalent action that can be used to construct an inequivalent
prequantisation of $\Gamma$.

Fractional statistics of particles were postulated for motion of particles in two
spatial dimensions in the presence of a solenoid with flux $\Phi$ \cite{Wilc}.
 We can crudely substitute for
the effects of the solenoid by excising a point from the
configuration space of the particle making the configuration space and hence the phase
space non-simply connected.
In fact, the resulting phase space has
topology ${\bf R}^2 \times ({\bf R}^2 -\{0\}) = {\bf R}^3 \times
S^1$. The homotopy group is clearly ${\bf Z}$ and all possible
characters of the group are of the form $\chi_{\alpha}(n) =
e^{i \alpha n}$, for arbitrary values of ${\alpha} \in {\bf R}$.
Now, ${\bf R}^4 $ is the universal cover of $\Gamma$ and this has
a unique prequantisation with a trivial bundle. We shall denote it as $\tilde{\Gamma}$.

 Let us denote the
action of $\pi(\Gamma) = {\bf Z}$ on $\Gamma$ as $(n  \cdot (x,
I) \rightarrow (n\cdot x, I) $. Explicitly this equals
\begin{equation}
(x_1,x_2) \rightarrow (x_1 + n, x_2 +n)
\end{equation}
 These coordinates are not the homogeneous coordinates $(x,y)$
on ${\bf R}^2$ that correspond to the definition of the spatial
translations. The latter read in terms of them as $x = e^{(x_1+x_2)/2}
\cos \pi(x_1 - x_2), y = e^{(x_1 + x_2)/2} \sin \pi(x_1 - x_2)$. Effectively the action $n \cdot x$
corresponds to a rotation of $2 n \pi$ around the origin.

The possible actions of ${\bf Z}$ on the trivial bundle ${\bf R}^4
\times U(1)$ are then $n \cdot (x, I, e^{i \phi})
\rightarrow (n\cdot x, I , e^{i n \alpha} e^{i \phi})$. Each
of these actions gives a different bundle $Y_{\alpha}$ as a
quotient, this being a different prequantisation of $\Gamma$. This
fact is often referred to as implying the need of {\em multivalued
wave-functions} \cite{Wilc}. Clearly from the perspective of
geometric quantisation each choice of $\alpha$ is an {\em
intrinsic property} of the quantum system and defines a physically
different  Hilbert space, with wave functions corresponding to
cross-sections of different bundles.

 We should note that at
 least the spatial rotations (around the origin) are well defined
in our resulting phase space. Its action is in terms
of the coordinates $x$ and $y$: $ (x,y) \rightarrow (x \cos \phi +
y \sin \phi, - x
 \sin \phi + y \cos \phi)$. This action commutes with the action of ${\bf Z}$
 and is therefore well defined on $\Gamma$. A rotation of $2 \pi n$ is equivalent
 to a transformation $x \rightarrow n \cdot x$ and hence a phase change of
$e^{i n \alpha}$.

Consider now the combination of two such systems. The phase space has as universal cover
${\bf R}^4 \times {\bf R}^4$
which has a unique trivial prequantisation.  On this bundle we have the action of
${\bf Z} \times {\bf Z}$ as
\begin{equation}
(x_1, x_2, I_1, I_2, e^{i \phi}) = (n \cdot x_1, m \cdot x_2, I_1,
I_2, e^{i \alpha_1 n + i \alpha_2 m} e^{i \phi} )
\end{equation}

 What constitutes identity of particles
is now an issue, because of the nature of the parameters
$\alpha_1$ and $\alpha_2$ that have no classical analogue. From
the standard quantum treatment of such systems $\alpha$ depends on
the details of the experimental setup.  If the system
represents a solenoid with magnetic flux $\Phi$ through a plane,
then $\alpha =  q \Phi$, where $q$ is the charge of the
particles \cite{Wilc}.
Both $q$ and $\Phi$ are parameters characterising uniquely a given system (they are an intrinsic and
an extrinsic property respectively), hence
identity necessitates both
 particles to be characterised by the same value of $\alpha$. Now,
  the resulting phase space
for the system is obtained by excising the diagonal, i.e. the set
of all $x_1$ and $x_2$ such that $n \cdot x_1 = m \cdot x_2$ or
otherwise $(n-m) \cdot x_1 = x_2$. Transformations of the form
$(n,-n)$ do not preserve the diagonal and hence cannot be used to define
 $\Gamma \times \Gamma - \Delta$ as a quotient of $\tilde{\Gamma} \times \tilde{\Gamma} - \tilde{\Delta}$.
This implies that
 $\pi_1(\Gamma \times \Gamma - \Delta) = {\bf Z} $, a fact that
as can be also checked directly.

So on $\tilde{\Gamma}
\times \tilde{\Gamma} -\Delta$ we have a pullback of the bundle on
$\tilde{\Gamma} \times \tilde{\Gamma}$ and an action of ${\bf Z}$
of the form
\begin{equation}
(x_1, x_2, I_1, I_2, e^{i \phi}) =
 (n \cdot x_1, n \cdot x_2, I_1, I_2,
 e^{2 i \alpha n } e^{i \phi} ).
\end{equation}

Taking  the quotient with respect to this action we obtain the
bundle  that prequantises the combined system, before the
implementation of the exchange symmetry (this  bundle is what we
referred to as $i_*Y$ in section 3). The action of the permutation
group is established by the demand that two repeated exchanges
ought to give an identity. If  the exchange induces a phase change
$e^{i \theta}$ on the fibers one needs have $e^{2i \theta} = e^{2
i n \alpha}$ for some integer $n$. Then $e^{i \theta} = \pm e^{in
\alpha}$. But equation (B.4) states that  for even $n$ the action
of $e^{i n \alpha}$ is identical to that of unity. Hence  there
are only four distinct ways of implementing the phase change: the
standard two of Bose and Fermi statistics with multiplication in
the fibers by 1 and (-1) respectively, but also  two more
corresponding to multiplication by $+e^{i \alpha}$ and $- e^{i
\alpha}$ respectively.

This means that the permutation group acts as
\begin{equation}
([x_1, x_2, I_1, I_2, e^{i \phi}]_{\alpha} \rightarrow
[x_1, x_2, I_1, I_2, (-1)^f e^{i l \alpha} e^{i \phi})
]_{\alpha},
\end{equation}
where both $f$ and $l$ take values $0$ and $1$. The brackets
$[\ldots]_{\alpha}$ denote equivalence classes with respect to the
action (B.4).

For $\alpha$ an
integer multiple of $\pi$ we have only Bose and Fermi statistics. In this case the charge of
the particle is quantised  $ q  = n
\frac{\pi}{ \Phi}$.

If we  employ our principle in this phenomenological
system, we will notice that the condition that the smooth paths
implementing the rotation do not cross the diagonal again imply
that the total rotation $g_1 g_2$ (as explained in section 4.1)
has to take place an {\em odd} number of times $2 \pi$. This
corresponds to a phase change of $e^{ i \alpha}$. Hence,
of all possible statistics {\em the one characterised by $f=0$ and $l=
1$ is selected}. Note that for $\alpha = (2k +1) \pi $ this is
Fermi statistics, for $\alpha = 2k \pi $ it is Bose, in the
general case it is neither.

To summarise, the system we described admits generically
four different statistics for the combined systems
and only one of them is allowed by our version of the spin-statistics theorem. This
analysis is an illustration of how the framework of geometric quantization and our analysis can be employed
to deal with combination of identical  systems that admit action of a symmetry group and is not restricted
to the standard case of particle in three spatial dimensions.

\end{appendix}

\end{document}